\documentclass[modern]{aastex62}
\begin{document}

\title{The Leaky Pipeline for Postdocs: A study of the time between receiving a PhD and securing a faculty job for male and female astronomers}
\correspondingauthor{Kevin Flaherty}
\email{kmf4@williams.edu}

\author[0000-0003-2657-1314]{Kevin Flaherty}
\affil{Department of Astronomy and Department of Physics, Williams College \\
Williamstown, MA 01267, USA}

\begin{abstract}
The transition between receiving a PhD and securing a tenure track faculty position is challenging for nearly every astronomer interested in working in academia. Here we use a publicly available database of recently hired faculty (the Astrophysics Job Rumor Mill) to examine the amount of time astronomers typically spend in this transitory state. Using these data as a starting point to examine the experiences of astronomy postdocs, we find that the average time spent between receiving a PhD and being hired into a faculty position is 4.9$\pm$0.3 years, with female astronomers hired on average 4.2$\pm$0.4 years after receiving a PhD while male astronomers are typically hired after 5.3$\pm$0.4 years. Using a simple model of the labor market, we attempt to recreate this gendered difference in time spent as a postdoc. We can rule out the role of the increasing representation of women among astronomy PhDs, as well as any bias in favor of hiring female astronomers in response to efforts to diversify the faculty ranks. Instead the most likely explanation is that female astronomers are leaving the academic labor market, at a rate that is 3-4 times higher than male astronomers. This scenario explains the distinct hiring time distributions between male and female astronomers, as well as the measured percentage of female assistant professors, and the fraction of female applicants within a typical faculty search. These results provide evidence that more work needs to be done to support and retain female astronomers during the postdoctoral phase of their careers. 
\end{abstract}

\section{Introduction}

Across many of the factors that would make a faculty candidate appealing to a hiring committee, many studies have shown a persistent bias in favor of male scientists. Men are invited to give more colloquia than women \citep{nit17}, have a higher success rate at receiving telescope time \citep{rei14,lon16,pat16,spe18}, are cited more frequently \citep{cap17}, ask more questions at conferences \citep{dav14,pri14,sch17a,sch17}, are overrepresented among prestigious physics fellowships \citep{nor18}, and have their abstracts rated at a higher level of scientific quality \citep{kno13}. These biases exist despite the lack of evidence for biological differences in science and math ability \citep{hyd05,gig17}. Elite male faculty in the life sciences are less likely to take on female students \citep{she14}, limiting early entry points into the field, and recommendation letters show gender-based biases \citep{dut16,mad18}. Female scientists who are able to navigate these effects still have to deal with high rates of sexual harassment \citep{cla14,nat18}, and higher demands on their time in the form of an increased work load and more requests for favors \citep{el18}. These factors can lead to a female candidate being judged as less qualified than an otherwise equivalent male candidate. 


Even among candidates with identical qualifications there is a bias against women. \citet{mos12} found that identical resumes were graded differently based on the gender of the applicant; women were offered a lower salary and were less likely to be offered a position. More reservations are expressed about CVs from female faculty members \citep{ste99}, and in hiring decisions the relationship status of a candidate is more often discussed, always as a negative, for female applicants \citep{riv17}. 

Hiring biases based on race/ethnicity are also present \citep{ber04,gib16}, and astronomers at the intersection of multiple marginalized identities face increased levels of harassment \citep{cla17}. To our knowledge no study has been done on the hiring of scientists who identify as LGBTQI, although LGBTQI scientists face a challenging heteronormative environment \citep{cec11,ath16,yod16} and sexual minority students are less likely to persist in STEM majors \citep{hug18}.




Biases against hiring women can change the length of time it takes to transition from getting a PhD into a tenure-track faculty position. An otherwise qualified female candidate will need to send out more applications, potentially over a longer period of time, before securing a faculty position. Conversely, a push towards hiring women, in an effort to increase the diversity of the faculty ranks, may lead to women being hired more quickly out of graduate school. Here we examine the time between when an astronomer receives a PhD and when they are hired into a tenure track faculty position, using a publicly available listing of recent faculty hires\footnote{The data used in this paper, as well as the code used to create the figures and the model, are available at \url{https://github.com/kevin-flaherty/hiring_distro}. A Jupyter notebook with more detail on the analysis, and a more complete set of models, is available at \url{https://github.com/kevin-flaherty/hiring_distro/blob/master/hiring_demo.ipynb}}. In section~\ref{data} we discuss the data collection, and the finding that recently hired female assistant professors spent less time as a postdoc than male assistant professors. In section~\ref{model} we introduce a simulation of the labor market, and examine various models for recreating the gendered hiring time discrepancy, finding that a model in which female astronomers leave the labor market at a higher rate than male astronomers best explains the data. Finally we discuss the implications of these results, and possible methods for improving the support and retention of female astronomers. 

\section{Data Collection and Results\label{data}}

\subsection{Data}
The sample for this study is drawn from the Astrophysics Rumor Mill\footnote{\url{http://www.astrobetter.com/wiki/Rumor+Mill+Faculty-Staff}}. This webpage contains an editable wiki listing various job postings within astronomy, and is updated by astronomers during the job season to indicate e.g. when phone interviews are conducted, who is on the short list, and the name of someone that has accepted the position, if any of this information is known for a particular job. 

For the purposes of this study, it serves as a public listing of astronomers that have recently been hired into tenure-track faculty positions, with information on the year in which they were hired. While this is not a complete listing of everyone that has been hired into astronomy, it is a starting point for conducting such a study. For each rumor of an accepted job offer, which was subsequently confirmed through a Google search, we record the gender of the astronomer, the year that they were hired, the year they received a PhD, and whether they were hired by an R1 school or non-R1 school. The PhD year was determined based on a Google search for a CV, or other public listing of a graduation year, while the year that someone was hired was taken as the spring year of the hiring cycle (ie. for someone listed on the 2014-2015 rumor mill, we use 2015 as the year they were hired). We assume a gender binary (either male or female), based on information from a Google search (e.g. a picture, gendered pronouns, etc.)\footnote{We recognize the problematic nature of imposing a gender binary upon individuals that may not ascribe to a strict male/female gender. Follow up work must take better account of gender identity \citep[e.g.][]{tra16}.}. Data collection starts with the 2010-2011 job cycle and runs through the 2016-2017 cycle. We assume that all positions listed under the Tenure Track heading on the Astrophysics Rumor Mill are in fact tenure-track faculty positions, although we do not verify this assumption. 

We exclude astronomers that are moving between tenure-track positions since our research question is focused on the initial hiring decision. We also exclude anyone that received a PhD prior to 2000 since most of the astronomers in this range have spent a substantial amount of time working at a national lab (e.g. JPL, STSCI) before taking the tenure-track faculty position. Permanent positions at national labs are not included because of the small number of positions listed on the Rumor Mill, and the difference between the work environment at a national lab versus an academic institute. We also focus on colleges and universities in the US since hiring practices and job responsibilities can vary substantially between US and non-US institutions. Given the very small number statistics, and to avoid identifying particular individuals, we do not record race/ethnicity. From this list we exclude anyone for which we could not determine the gender or the year that they received their PhD. The Carnegie classification of the hiring institution is determined using the online lookup tool\footnote{\url{http://carnegieclassifications.iu.edu/lookup/lookup.php}}, and we only record if the institution is listed as R1 (Doctoral Universities - Highest research activity) or non-R1 (including institutions that award masters degrees, as well as undergraduate only institutions). 

This leaves a total sample of 245 astronomers, 157 of which are men and 88 of which are women. The fraction of women within our sample (35$\pm$4\%) is slightly higher than the fraction of female assistant professors in 2013 (26$\pm$4\%, \citet{hug14}) suggesting that our sample is over-represented in terms of female astronomers. As discussed below, this slight bias is unlikely to be responsible for the  discrepancy in hiring time between male and female astronomers. We do not find a significant difference in the hiring time from one job cycle to the next, and consider the entire sample as a whole in order to minimize Poisson errors. 

\subsection{Results}
\begin{figure}[!htb]
\center
\includegraphics[scale=.46]{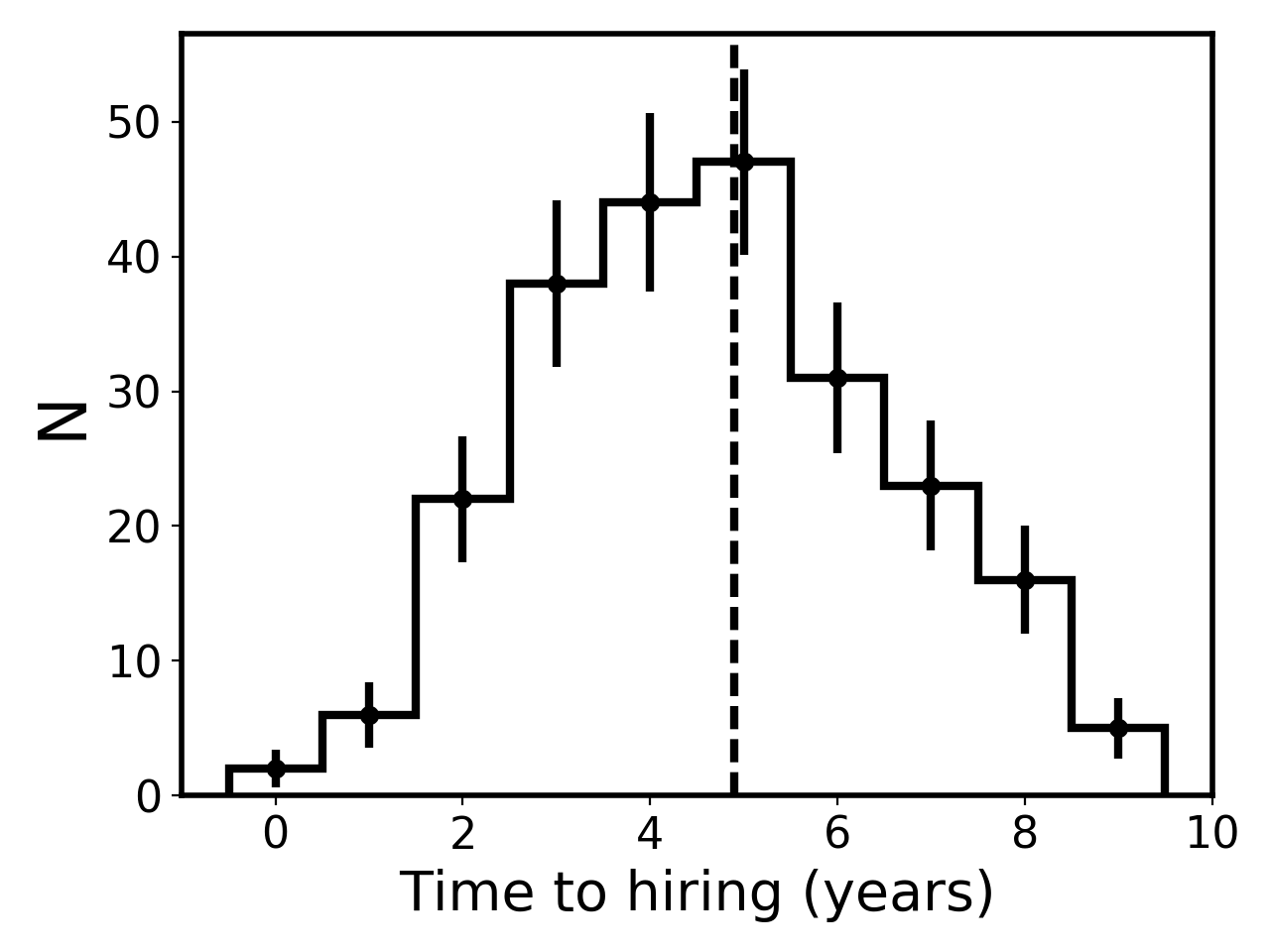}
\includegraphics[scale=.46]{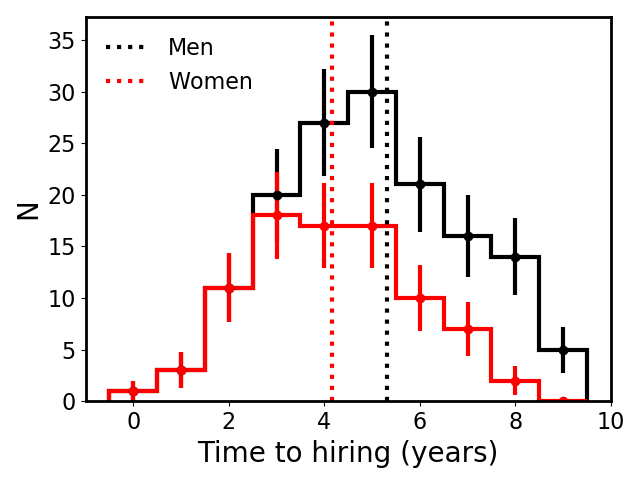}
\caption{(Left:) Number of astronomers hired into tenure track faculty positions, as a function of time since receiving their PhD. On average astronomers spend 4.9$\pm$0.3 years in a postdoc position. 
(Right:) Splitting the sample between male (black line) and female (red line) astronomers, we find a significant difference in the hiring time distribution.  This leads to female astronomers being hired, on average, one year earlier than male astronomers (4.2$\pm$0.4 vs 5.3$\pm$0.4).\label{hiring_time}}
\end{figure}

Figure~\ref{hiring_time} shows the distribution of time between receiving a PhD and being hired into a tenure track faculty position based on our sample. We find that, on average, astronomers spend 4.9$\pm$0.3 years in postdoc positions before being hired. Note that we only record when someone is hired, and not when they actually start their faculty job. Astronomers hired very soon out of graduate school may defer the start of their faculty position for a year or two, and as a result our measured mean hiring time is likely a lower limit on the average time astronomers spend as a postdoc. We do not find any significant difference in the hiring time between astronomers hired into R1 institutions (4.9$\pm$0.4 years) and non-R1 institutions (4.8$\pm$0.6 years). 

Figure~\ref{hiring_time} also shows the hiring time split based on gender, where we find a significant difference in how long it takes male and female astronomers to be hired into tenure track positions. The KS probability of these two distributions being drawn from the same population is only 0.03, with female astronomers hired on average 4.2$\pm$0.4 years after receiving their PhD, while for male astronomers the average hiring time is 5.3$\pm$0.4 years. This difference can be seen in the shape of the hiring time distributions. Within the first two years out of graduate school, the number of male and female astronomers hired into faculty positions is identical, while from years four through ten twice as many male astronomers as female astronomers are hired. Note that women are slightly over-represented in this sample, implying that among all astronomers {\it at least} twice as many male astronomers as female astronomers are hired in years four through ten. There is no significant evidence that the hiring time difference between male and female astronomers depends on the Carnegie classification of the hiring institution, although the small number of astronomers hired by non-R1 schools limits this conclusion. We do find that among R1 hires 68$\pm$6\% (124 out of 182) are men, while among non-R1 hires 52$\pm$19\% (33 out of 63) are men.

\section{Faculty Hiring Model\label{model}}
To understand the source of the discrepancy between when male and female astronomers are hired into tenure track faculty positions, we create a model of the hiring process, and introduce gendered factors in an attempt to reproduce the hiring time distribution for women.

The basic model consists of $N_{\rm phd}$ astronomers added to the labor pool each year, and $N_{\rm hire}$ astronomers randomly selected from the full labor market and `hired' into a faculty position each year. Once an astronomer has been hired they are removed from the labor market, and the model proceeds to the next year adding another $N_{\rm phd}$ astronomers to the labor pool and removing another $N_{\rm hire}$ astronomers from the labor pool. Whenever someone is hired we record the year they were hired, the year they received their PhD, and their gender. Unless otherwise specified, we assume that 30\% of the $N_{\rm phd}$ astronomers receiving a PhD and entering the labor market are women. 

To reproduce the non-uniform hiring time distribution we assume that the probability of being hired, $p_{\rm hire}$, depends on the time since the astronomer received their PhD, $t$. For $t>10$ years we assume $p_{\rm hire}$=0, while the probabilities for $t<10$ are manually adjusted. The $p_{hire}$ term is the product of the probability of an astronomer applying for a faculty job, and the probability that they will subsequently be hired. For the purposes of this study we do not distinguish between these factors, although they likely change with time (e.g. very few astronomers apply to faculty jobs directly out of graduate school).

The labor pool is populated starting in 1980, while hiring begins in 1990. As with the data, we only consider astronomers that were hired between 2011 and 2018. The model is started well before 2011 to ensure that it has reached a steady state before comparing with the data.

To avoid large statistical noise in the model, we use values of $N_{\rm phd}$ and $N_{\rm hire}$ that are much larger than reality, but still preserve the ratio of tenure track jobs to PhD astronomers. \citet{mul14} report $\sim$150 PhDs were awarded in astronomy from 2007 to 2012, after averaging $\sim$110 in the prior decade\footnote{AIP maintains enrollment and degree records going back to 1978 \url{https://www.aip.org/statistics/rosters/astronomy}}, while in the Astrophysics Rumor Mill there are roughly 50 tenure track faculty positions listed each year. In the model we use $N_{\rm phd}$=30000 and $N_{\rm hire}$ = 10000, consistent with the 3:1 ratio of PhDs awarded to faculty jobs available each year. This is may be an over-estimate of the $N_{\rm phd}$/$N_{\rm hire}$ ratio since we do not account for non-US astronomers that apply to US faculty jobs, and some of the US faculty positions listed on the Astrophysics Rumor Mill are for physics positions for which an astronomer is not hired, although we do assume that all astronomers that receive a PhD enter the faculty job market. Our results will not be strongly affected by this discrepancy, unless $N_{\rm phd}$ is nearly identical to $N_{\rm hire}$ which is highly unlikely. To compare the results of the model with the data, we normalize each hiring time distribution by the total number of hired male/female astronomers and compare the relative fraction of male/female astronomers hired as a function of time. 

We consider three explanations for the difference between the hiring time distributions of male and female astronomers: (1) An increase in the number of women receiving astronomy PhDs over time, (2) A higher probability of a female astronomer being hired relative to a male astronomer, (3) Women leaving the faculty labor market at a higher rate than male astronomers. The values of $p_{hire}(t)$ are set based on the male hiring time distribution, while the scenario-specific parameters (e.g. the rate at which female astronomers leave the labor market) are adjusted to best reproduce the hiring time distribution for women. 

We also compare our model predictions with the general demographics of the labor market. Specifically we anchor the fraction of women entering the labor market to the fraction of female PhD students in 2003 (30$\pm$2\%), compare the resulting pool of hired astronomers to the fraction of female assistant professors in 2013 (26$\pm$4\%) as measured by \citet{hug14}, and compare the fraction of women in the entire applicant pool to the fraction of women in the applicant pool of tenure track faculty positions (19$\pm$3\%) as measured by \citet{tho15}. To calculate the number of female PhD students in 2003 we count all entries that received a PhD between 2003 and 2010. This range of time encompasses first year students and students that have recently graduated, assuming a typical PhD length of 7 years. To calculate the fraction of female assistant professors in 2013, we measure the fraction of women among astronomers hired between 2007 and 2013 within our model. This assumes a typical time to tenure, and promotion from assistant to associate faculty, of six years. To compare with the findings of \citet{tho15} on the fraction of women among applicants to tenure track positions, we calculate the fraction of women within the labor pool in 2014. 

\subsection{Changing PhD Demographics}
The first model we consider is one in which the fraction of women within astronomy increases with time. Between 1992 and 2013 the fraction of female graduate students rose from 22\% to 34\%, while the fraction of female assistant professors increased from 17\% to 26\% \citep{hug14}. Adding substantially more young women to the labor pool will bias the hiring time distribution towards shorter times, if the increase in female representation is steep enough.

\begin{figure}[!hbt]
\center
\includegraphics[scale=.6]{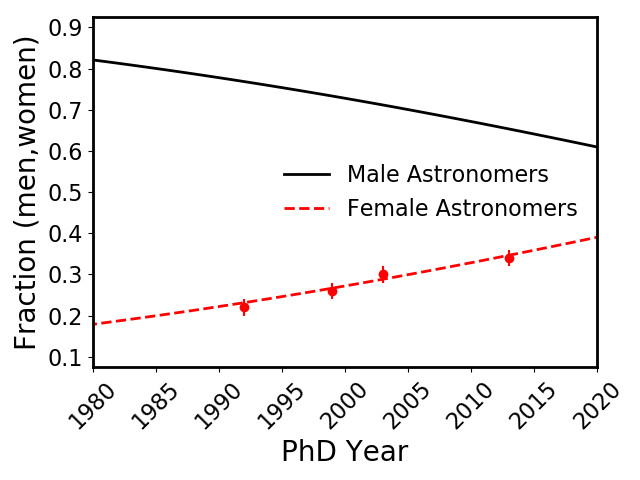}
\caption{Fraction of astronomy PhDs being awarded to men (black line) and women (red dashed line) as a function of time as predicted by our model prescription, compared to the measurements of \citet{hug14} (red points).\label{f_phd}}
\end{figure}

We model the change in the fraction of astronomy PhDs going to women as a logistic function
\begin{equation}
f_{\rm phd} = \frac{1}{1+\exp (-s(t-t_0))}\\
\end{equation}
where $s$ controls the rate of increase in the fraction of women, $t_0$ is the year in which gender parity is reached, and $t$ is the year. We find that $s$=0.027 and $t_0$=2040 provide a reasonable fit to the demographics survey of \citet{hug14} (Figure~\ref{f_phd}). In this model we assume that $p_{hire}(t)$ is identical between male and female astronomers, with the only difference arising from the increasing number of women in the labor market with time. 

\begin{figure}[!hbt]
\center
\includegraphics[scale=.6]{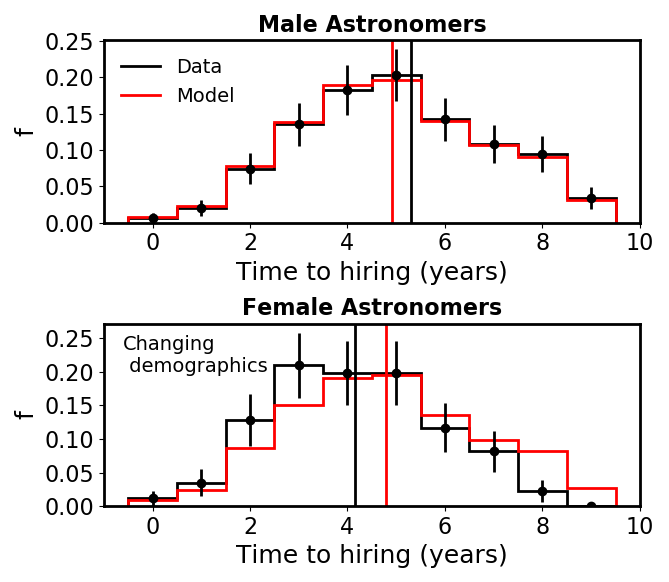}
\caption{Relative fraction of astronomers hired as a function of year since receiving their PhD, split based on gender. The data are shown in black, with Poisson uncertainties, while a model based on the increasing fraction of PhDs received by female astronomers with time is shown in red. The relative probability of being hired in a given year ($p_{\rm hire}(t)$) is fit to the male hiring time distribution, and assumed to be the same for female astronomers, with the only difference arising from the increased number of female astronomers within the labor market with time. This model is unable to reproduce the hiring time distribution for female astronomers, suggesting that a change in demographics is not responsible for the shift toward shorter postdocs for women.  \label{model1}}
\end{figure}

Figure~\ref{model1} compares the predicted male and female hiring time distributions for this model to the measured distributions. By construction this model reproduces the male hiring time distribution, but the increased fraction of women receiving PhDs over time does not substantially shift the female hiring time distribution toward shorter timescales. This model is able to reproduce the fraction of female assistant professors in 2013 (28\% vs 26$\pm$4\%) but it over-predicts the fraction of women in the labor market (24\% vs 19$\pm$3\%). The inability of this model to reproduce the hiring time distribution for female astronomers, and the over-prediction of women in the labor market, indicates that another effect is needed to explain the data. Because of the small influence of changing demographics on the model results we do not include this correction in subsequent models. 

\subsection{Bias Toward Hiring Women}
In our second model, we consider a bias in favor of hiring women in tenure track faculty positions. This may occur as a result of diversity efforts leading more colleges and universities to hire women more quickly out of graduate school, or because women are intrinsically better qualified for faculty jobs. Given the finite number of women in the labor market, by quickly removing women there will be fewer female astronomers available at later times, shifting the peak of the hiring distribution towards shorter timescales.

We model a bias towards female astronomers by introducing a parameter $b$ that is a multiplicative increase in the probability of hiring a female astronomer relative to hiring a male astronomer (e.g. $b=2$ indicates that a female astronomer is twice as likely to be hired as a male astronomer). This multiplicative factor is applied to all female astronomers regardless of when they receive their PhD. 

\begin{figure}[!hbt]
\center
\includegraphics[scale=.4]{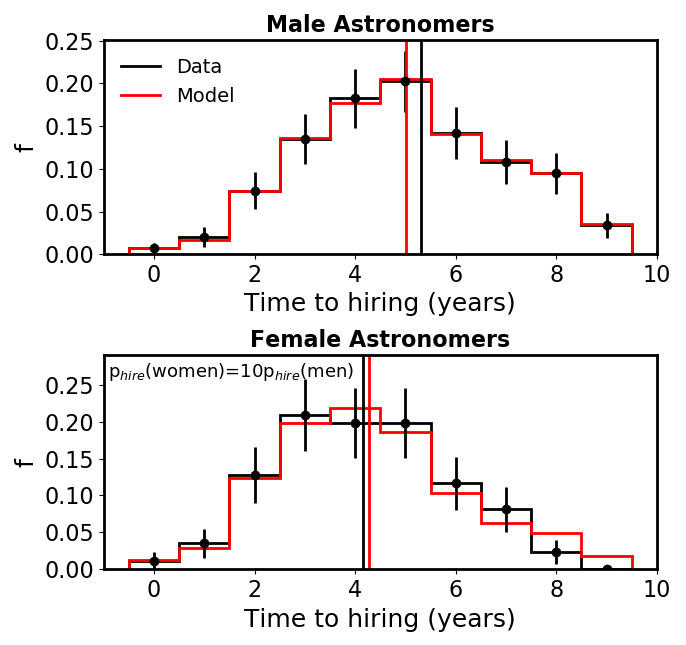}
\includegraphics[scale=.4]{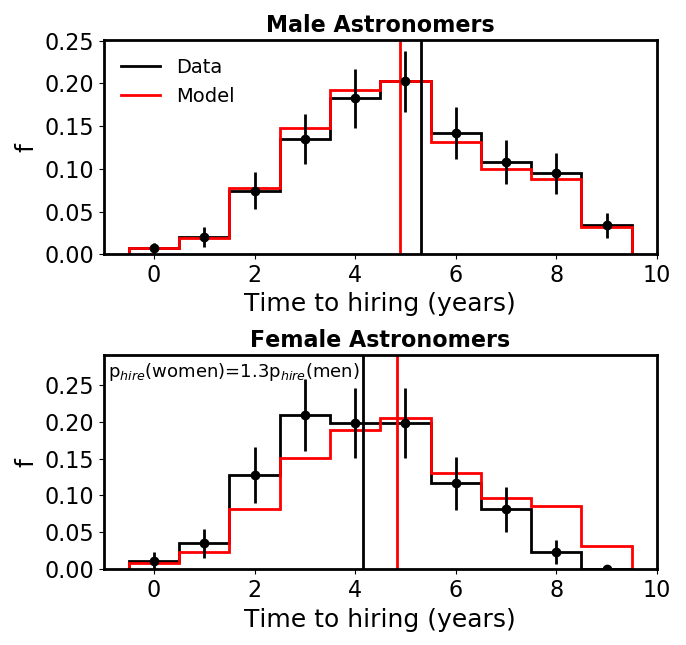}
\caption{Similar to Figure~\ref{model1}, except for a model in which the probability of a female astronomer being hired is a factor of $b$ times larger than for male astronomers. We can reproduce the hiring time distribution for female astronomers only if we assume that female astronomers are 10 times more likely to be hired into a faculty position (left panel). We can rule out this model based on the fact that it predicts that 70\%\ of assistant professors are female, when in reality the fraction is only 26$\pm$4\% \citep{hug14}. A model (right panel) that is consistent, within 3$\sigma$, of the measured fraction of female assistant professors ($b$=1.3, predicting 34\%\ of assistant professors in 2013 are women) is unable to reproduce the hiring time distribution for women. \label{model2_b10}}
\end{figure}

We are able to reproduce the normalized hiring time distribution for female astronomers (Figure~\ref{model2_b10}), but only with a very strong bias in favor of hiring female astronomers ($b=10$). It is highly unlikely that a female astronomer is 10 times more likely to be hired than a male astronomer, especially given that this predicts that 70\%\ of assistant professors in 2013 are female, when in reality only 26$\pm$4\% are female. A 30\%\ bias ($b$=1.3) results in a fraction of female assistant professors (34\%) that is consistent within 3$\sigma$ of the measured value, but is unable to reproduce the hiring time distribution for women (right panel of Figure~\ref{model2_b10}) 

The need for a strong bias in this scenario is because the number of PhDs produced each year far outnumbers the number of tenure track faculty jobs available, even without accounting for non-US PhD students entering the US labor market (which will increase $N_{\rm phd}$), or the presence of physics positions on the the rumor mill (which will reduce $N_{\rm hire}$). Under these conditions it takes a substantial bias in favor of women in order for the number of female astronomers hired each year to substantially deplete the labor market. Based on the hiring time distribution and the fraction of female assistant professors, we can rule out a substantial bias in favor of hiring female astronomers as a plausible model. 

\subsection{Women Leaving the Labor Market}
The third model we consider is one in which astronomers leave the labor pool at a rate that increases with time. There are many factors that can drive someone out of the academic job market and these factors may differ between male and female astronomers.

To model this effect we introduce an exponential function that defines the fraction of the astronomers leaving the labor market as a function of time since receiving a PhD.
\begin{equation}
f_{\rm leave} = 1-\exp (-(t-2)/\tau).
\end{equation}
We assume that no one leaves until two years after receiving a PhD, and that the shape of the subsequent function is defined by the variable $\tau$. As discussed below, allowing for some fraction of astronomers to leave directly out of graduate school produces similar results. A smaller $\tau$ implies that the fraction of astronomers leaving the faculty labor market increases rapidly with time, while a larger $\tau$ results in a smaller fraction of astronomers leaving the labor market. Female and male astronomers have different values of $\tau$ to allow for different departure rates. The probability of being hired is assumed to be identical between male and female astronomers, and the only gendered difference is in the number of astronomers available to be hired at a given time. 

\begin{figure}[!hbt]
\center
\includegraphics[scale=.6]{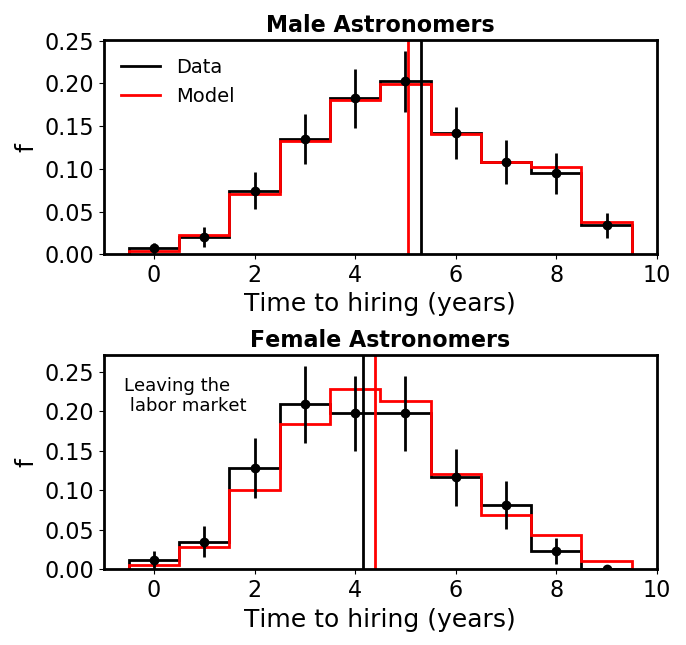}
\caption{Similar to Figure~\ref{model1}, except for a model in which male and female astronomers leave the faculty labor market. The fraction of astronomers leaving the labor market increases with time, and is 3-4 times higher for female astronomers than for male astronomers. This model is able to reproduce the hiring time distribution for female astronomers and also predicts that women make up 18\%\ of the labor market and 24\% of assistant professors in 2013, similar to the measured fractions (19$\pm$3\% and 26$\pm$4\% respectively).  \label{model3}}
\end{figure}

As shown in Figure~\ref{model3}, this model is able to produce a reasonable match to the normalized hiring time distribution for female astronomers with $\tau_{\rm male}$=40 and $\tau_{\rm female}$=10. It also predicts that women make up 18\% of the labor market and 24\% of assistant professors in 2013, consistent with the measured values of 19$\pm$3\% \citep{tho15} and 26$\pm$4\% \citep{hug14} respectively. The ability to reproduce the hiring time distribution, the fraction of female assistant professors, and the fraction of women applying to faculty jobs provides strong support for the presence of the leaky pipeline for women at the postdoctoral level. 

\begin{figure}[!hbt]
\center
\includegraphics[scale=.6]{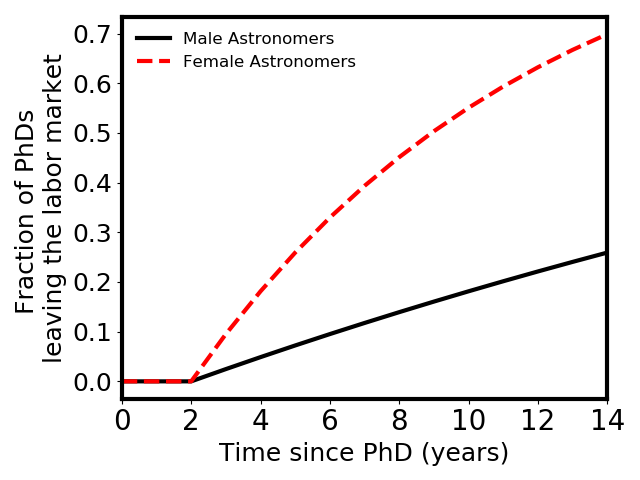}
\caption{The fraction of PhDs astronomers leaving the academic labor market as a function of time since receiving a PhD for the model that is best able to reproduce the hiring time distributions. Within this model the fraction of women leaving the labor market is 3-4 times higher than the fraction of male astronomers leaving the labor market. \label{df_model3}}
\end{figure}

Within this model the fraction of women leaving the labor market is 3-4 times higher than the fraction of men leaving the labor market (Figure~\ref{df_model3}). Given the relative proportion of female and male astronomers entering the labor market, this means that, in terms of total number, as many female astronomers (if not more) leave the academic labor market as male astronomers, despite the fact that male astronomers strongly outnumber female astronomers. 

\begin{figure}[!hbt]
\center
\includegraphics[scale=.4]{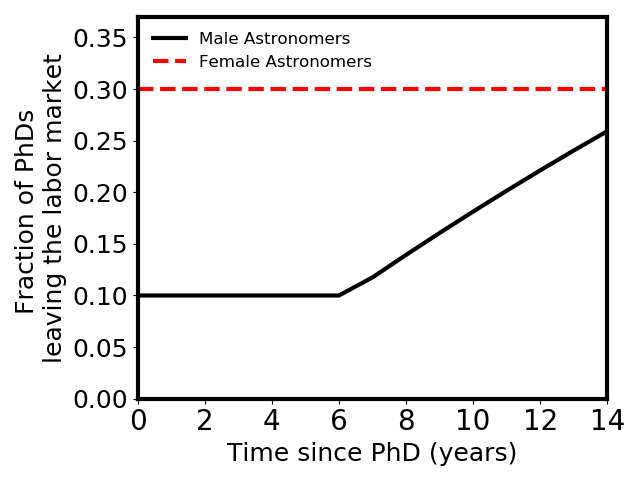}
\includegraphics[scale=.45]{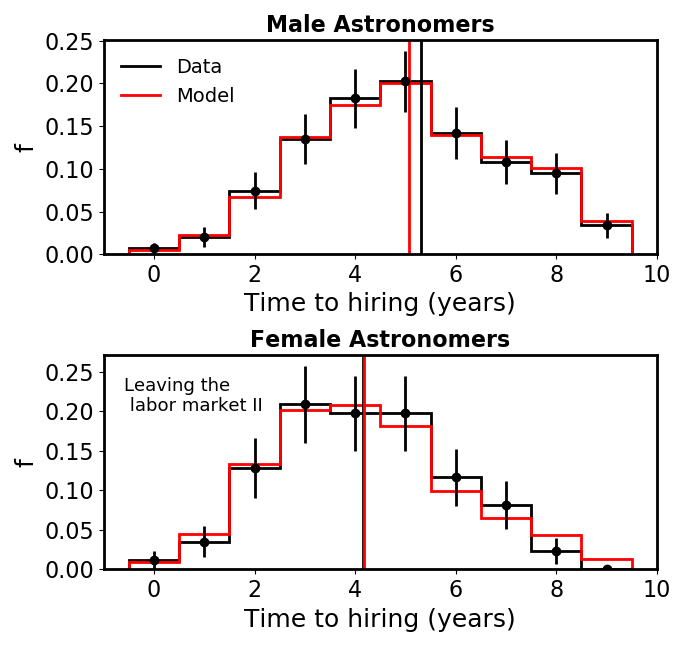}
\caption{A model with a different functional form for the departure rate of male and female astronomers from the faculty labor market can still match the hiring time distribution. (Left:) We consider a model in which there is a floor to the fraction of astronomers leaving the field, with a higher floor for women than for men. (Right:) Similar to the model in Figure~\ref{model3} in which no astronomers leave straight out of graduate school, this model is able to produce a reasonable match to the hiring time distribution for female astronomers. This highlights the degeneracies present in the model, but highlights the need for the departure rate of women to be higher than for men to match the hiring time distribution.\label{model3e}}
\end{figure}

This model is not a unique match to the available data, which can be demonstrated by including a floor to the fraction of male/female astronomers leaving each year (Figure~\ref{model3e}). Here we assume that at least 10\%\ of all male astronomers depart the faculty labor market each year out of graduate school, while at least 30\%\ of female astronomers depart each year.  This model is able to match the hiring time distribution for female astronomers despite the change in the exact rate at which women leave the faculty labor market from the model shown in Figure~\ref{model3}. This new model also predicts that women make up 20\%\ of the labor market in 2014, and 18\%\ of assistant professors in 2013, consistent with the measured values (19$\pm$3\%\ and 26$\pm$4\%\ respectively). The key feature of the models shown in Figures~\ref{model3} and~\ref{model3e} is that women leave the labor market at a higher rate than men. While the data are not detailed enough to constrain the exact functional form of this difference, a difference in the departure rate must be present to reproduce the data.  

\section{Conclusions and Ways Forward}
The postdoctoral stage is an important phase of nearly every astronomer's career. Often viewed as some of the most productive years in terms of research output, they can also be some of the most stressful years given the transitory nature of postdoctoral jobs. A large portion of a postdoc's time is spent searching, applying, and interviewing for long term positions. Here we find that among recently hired astronomy faculty this phase lasted on average 4.9$\pm$0.3 years. We also find that female assistant professors spent an average of 4.2$\pm$0.4 years as a postdoc while male assistant professors spent an average of 5.3$\pm$0.4 years as a postdoc. Through a model of the labor market we find that this gendered difference is not due to a bias in favor of hiring female astronomers in response to efforts to increase the diversity of the faculty ranks, but is most likely due to women leaving the academic labor market at a higher rate than their male counterparts. While there are limitations to our sample, this study is a starting point for examining the role of gender in defining the experiences of astronomy postdocs. 

This analysis indicates that the combined effect of the structural barriers outlined in the introduction is to drive otherwise qualified astronomers away from academic jobs. What can be done to improve upon the support and retention of female astronomy postdocs? A number of recent guides have been developed with detailed suggestions for the support of, and the removal of systematic barriers faced by, female and minority scientists\footnote{Recent examples include the \href{https://tiki.aas.org/tiki-index.php?page=Inclusive_Astronomy_The_Nashville_Recommendations}{Inclusive Astronomy Recommendations},  \href{https://arxiv.org/abs/1804.08406}{LGBT+ Inclusivity in Physics and Astronomy: A Best Practices Guide}, Jarita Holbrook's \href{https://arxiv.org/abs/1204.0247}{Survival Strategies for African American Astronomers}, the American Physical Society's \href{https://www.aps.org/programs/women/reports/cswppractices/index.cfm}{Effective Practices for Recruiting and Retaining Women in Physics}, and the National Academies of Science, Engineering, and Medicine's \href{https://doi.org/10.17226/24994}{report on sexual harassment}}. Two common themes among these guides are the importance of good mentorship and the challenges associated with the two body problem. The importance of these two factors is highlighted by the longitudinal study of graduate student experiences presented in \citet{ivi16}. They find that while gender is not a direct factor in leaving the field, it is strongly correlated with a negative advisor relationship, and a higher frequency of reporting two body problems, both of which are directly responsible for astronomers leaving the field. \citet{mcc18}, in a national survey of 7600 postdocs across multiple disciplines, also find that career choice is strongly influenced by the mentorship that a postdoc receives from their supervisor. 

What does good mentorship look like at the postdoctoral stage? The APS guide for supporting postdoctoral researchers\footnote{\url{https://www.aps.org/programs/women/reports/cswppractices/postdoctoral.cfm}} offers some concrete suggestions, including that postdocs should be encouraged to network outside of their research group, to broaden their CV beyond simply writing papers, and to have a well-developed career path. These suggestions reflect the fact that a postdoctoral position is a stepping stone to a more permanent job, and that scientists need additional training beyond what they received in graduate school in order to be fully prepared for these future jobs, regardless of whether or not these jobs are in academia. \citet{mcc18} find that feelings of career preparedness, in addition to support from the mentor with regards to the postdoc's chosen career plan, are significant factors in postdoc satisfaction. 

The two-body problem represents a different set of challenges. The term `two-body problem' often refers to the difficulty of finding geographically nearby jobs for two partners within a relationship, a difficulty that is exacerbated during the postdoctoral phase of a scientists career given the multiple job/location changes that are involved. Added on top of this is the possibility of having children, amidst the changes in location and in health/child care coverage. What can be done to minimize complications associated with the two-body problem and/or having children? One possibility is lengthening the typical duration of a postdoc from 3 years to e.g. 5 years. The analysis above indicates that most astronomers would need only one 5-year position, rather than two 3-year positions, before moving into a tenure-track faculty job. Longer postdoc positions would reduce the number of job transitions, and hence reduce the probability of running into complications associate with the two-body problem. Any scientist (either male or female, in dual-sex or single-sex relationships) that decides to have children (either through birth or adoption) during a postdoc will be subject to a more consistent set of health care and child care coverage if they are employed in one long-term postdoc position rather than multiple short-term postdoc positions. Longer postdoc positions are also better able to accommodate parental leave (for both parents, and for birth or adoption). Better support during parental leave for postdocs would also be valuable; the Inclusive Astronomy recommendations suggest three to six months of paid parental leave, available to both parents. Providing parental leave for adoption, in addition to birth, also provides support for same-sex couples \citep{ath16}, as well as dual-sex couples and individuals that are not in a couple that do not pursue child birth.

In addition to a positive mentor relationship and better support of the two-body problem, addressing the systematic barriers outlined in the introduction would certainly help improve the retention of female postdocs. This includes being aware of the biases present in all people when evaluating the work of otherwise equivalent scientists \citep{ste99,mos12,riv17}, and of the biases that prevent otherwise equivalent scientists from having identical CVs (e.g. lower success rates on telescope proposals). Online tools can search for biased language in letters of recommendation\footnote{\url{https://www.tomforth.co.uk/genderbias/}}, and there are also online guides on methods to mitigate biases in hiring decisions\footnote{\url{http://advance.umich.edu/strideResources.php}, \url{http://aliciaaarnio.solar/stellar/files/Hiringpracticesinastronomy.pdf}}. Reducing sexual harassment, which has been experienced by 28\%\ of all postdocs with 88\%\ of victims being women \citep{sle17}, is another significant factor that will improve the retention of female astronomers in academia. Recently the \citet{nat18} highlighted the negative effects of gender harassment (e.g. sexist jokes or the use of sexually crude terms) which they found to most prevalent in organizations that tolerated such behavior. While changing the culture within an academic department can be challenging, the cost of inaction is much higher.

While we have focused on astronomers hired into tenure track positions, this is not meant to suggest that a tenure track faculty position is the preferred career track for everyone that receives a PhD in astronomy. Nevertheless, many of the suggestions outlined above will help astronomers, and scientists in general, on all career paths. Positive mentoring and support in the face of the two body problem will help scientists beyond the postdoctoral phase of their career while reducing sexual harassment is a major challenge across all scientific disciplines, at all career stages. Here we have highlighted one aspect of the leaky pipeline as it applies to postdocs in the hopes that it will guide future efforts in reducing the barriers facing female and minority scientists.

\acknowledgements
I want to thank the Wesleyan STEM Diversity Journal Club for their thoughts about the results, and for providing an opportunity to discuss these issues among a supportive group. I also want to thank Johanna Teske for sharing her detailed thoughts on an early draft of this paper.

\end{document}